\documentclass[
 reprint,
 amsmath,amssymb,
 aps,prl,
 superscriptaddress,
 nofootinbib
]{revtex4-1}

\usepackage{graphicx}
\usepackage{dcolumn}
\usepackage{bm}
\usepackage{mathrsfs}
\usepackage{booktabs}
\usepackage{wasysym}

\begin{document}

\preprint{APS/123-QED}

\title{CMB Bispectrum from Non-linear Effects during Recombination}

\author{S.-C.~Su}
\affiliation{Centre for Theoretical Cosmology, Department of Applied Mathematics and Theoretical Physics,\\
University of Cambridge, Wilberforce Road, Cambridge CB3 0WA, United Kingdom}
\author{Eugene A.~Lim}
\affiliation{Centre for Theoretical Cosmology, Department of Applied Mathematics and Theoretical Physics,\\
University of Cambridge, Wilberforce Road, Cambridge CB3 0WA, United Kingdom}
\affiliation{Theoretical Particle Physics and Cosmology Group,
Physics Department,\\ Kings College London, Strand, London WC2R 2LS, United Kingdom}
\author{E.~P.~S.~Shellard}
\affiliation{Centre for Theoretical Cosmology, Department of Applied Mathematics and Theoretical Physics,\\
University of Cambridge, Wilberforce Road, Cambridge CB3 0WA, United Kingdom}

\date{\today}

\begin{abstract}
We study cosmological perturbations by solving the governing Boltzmann and Einstein Field equations up to second order, and calculate the corresponding CMB bispectrum during recombination. We include all the second-order Liouville and collision terms, truncating the multipole hierarchy at $l=10$, consistently including all $m\neq 0$ terms when calculating the bispectrum in the flat-sky limit.   At this stage, we focus on contributions at recombination and we neglect 2nd-order vector and tensor terms, lensing effects, and late time non-linear ISW. We find that the signal-to-noise for the bispectrum is 0.69 for $l_{max}=2000$, yielding an overall signal $F_{\rm NL} = 3.19$ (normalised relative to the local model). We find that the effective $f_{\text{NL}}$'s of the equilateral and local type are $5.11$ and $0.88$ respectively.  This recombination signal will have to be taken into account in a quantitative analysis of the Planck data. 

\begin{description}
\item[PACS numbers]
98.80.-k, 98.80.Es
\end{description}
\end{abstract}

\maketitle

\emph{\textbf{Introduction}.} With the Planck data shortly to become available, we may be able to detect non-Gaussianities in the Cosmic Microwave Background(CMB) for the very first time. Such a detection could provide crucial insights advancing our understanding of the physics of the very early Universe, so its importance cannot be overstated.  However, as well as primordial effects when the initial conditions are laid down during inflation, there can be late-time non-linear interactions of the photon-baryon fluid with gravity(For more details, refer to \cite{NGReview}). There have been several past attempts at estimating the CMB bispectrum from the latter in the literature \cite{TempDef,NBispec2,PitrouBispec}. A detailed quantitative analysis of these late-time contributions is necessary, not only to debiase potential primordial signals, but also as an independent test of GR and cosmological perturbation theory. In this paper, we present our numerical implementation of the calculation of the bispectrum produced at recombination and beyond. We obtain $f_{NL}^{local}=0.88$, in agreement with previous work focussing on the squeezed limit \cite{AnalyticSqu,AnalyticSqu2}. We also find an equilateral signal with $f_{NL}^{equil}=5.11$ which agrees very well with \cite{EquilFNL}.  The overall signal-to-noise $(S/N)_{\rm rec}=0.69$ provides an effective $F_{\rm NL} = 3.19$ contribution which is larger than anticipated previously and should be incorporated in the analysis of the Planck CMB data with a forecast variance of  $\Delta F_{\rm NL} = 5$.  First, however, we briefly describe the underlying analytic methodology before describing the numerical pipeline that has been developed, leading to the recombination bispectrum results.  A much more detailed discussion of these methods and results will be presented in a longer paper shortly \cite{SLS}.   

\smallskip
\emph{\textbf{Second-Order Perturbations and their Bispectra}.}
To study non-Gaussianities generated by non-linear effects, we solve the 2nd-order Boltzmann Equations
\begin{eqnarray}\label{Boltzmann}
\mathfrak{L}^{\text{[II]}}+\mathfrak{L}^{\text{[I,I]}}
&=&\mathfrak{C}^{\text{[II]}}+\mathfrak{C}^{\text{[I,I]}}
\end{eqnarray}
where $\mathfrak{L}$ is the Liouville operator, $\mathfrak{C}$ is the collision operator, the superscripts [II] and [I,I] denote linear terms of purely second-order perturbations and cross terms of two first-order perturbations respectively. In addition, we require knowledge of the first-order ionization fraction \cite{1stOpacity} at recombination, which encodes the effect of the perturbed last-scattering surface(LSS), when evaluating $\mathfrak{C}^{\text{[I,I]}}$.

We begin with the metric defined up to second-order
\begin{gather}\label{Metric}
-\frac{g_{\mu\nu}}{a(\tau)^2}dx^{\mu}dx^{\nu}= (1+2\Phi)d\tau^2-2\mathscr{B}_i dx^id\tau
\nonumber \\
-[(1-2\Psi)\delta_{ij}+2\mathscr{X}_{ij}]dx^i dx^j
\end{gather}
where $a(\tau)$ is the scale factor, and $\Phi$, $\Psi$, $\mathscr{B}_i$'s and $\mathscr{X}_{ij}$'s are perturbations expanded in the following way
\begin{eqnarray}\label{PhiPert}
\mathcal{A} = \mathcal{A}^{\text{[I]}} + \frac{1}{2}\mathcal{A}^{\text{[II]}}
\end{eqnarray}
with $\mathcal{A}$ denoting the perturbation quantities above. We will use Newtonian gauge up to second-order throughout. Moreover, we ignore the 1st-order and 2nd-order vector and tensor perturbations, i.e.~$\mathscr{B}_i^{\text{[I]}} = \mathscr{B}_i^{\text{[II]}} = 0$ and $\mathscr{X}_{ij}^{\text{[I]}} = \mathscr{X}_{ij}^{\text{[II]}} = 0$.\footnote{In general, the second-order vector and tensor perturbations are non-negligible as they are sourced by the first-order scalar perturbations nonlinearly, however, we can expect these to be subdominant.}

We compute the corresponding reduced bispectrum during recombination in the flat-sky and thin-shell approximation \cite{PitrouBispec}
\begin{eqnarray}\label{Bispectrum}
&b&_{l_1l_2l_3}\approx \frac{r_{\text{LSS}}^{-4}}{(2\pi)^2}\int^{\infty}_{-\infty}dk^z_1dk^z_2 P(k_1)P(k_2)
\int^0_{r_{\text{LSS}}}dr_1 dr_2 dr_3\nonumber\\
&e&^{-i(k_1^z r_1+k_2^z r_2+k_3^z r_3)}
S^{\text{[I]}}(k_1,r_1)S^{\text{[I]}}(k_2,r_2)
S_{\text{2ND}}(\bold{k}_1,\bold{k}_2,r_3)\nonumber\\
&~&+1\leftrightarrow 3 + 2\leftrightarrow 3
\end{eqnarray}
where LSS denoting the last-scattering surface, $r\equiv \tau_0 - \tau$ with the present conformal time $\tau_0$, $P(k)$ being the primordial power spectrum, $k^z$ being the component of $\bold{k}$ perpendicular to the tangent plane \cite{FlatSky}. 
The first-order source function $S^{\text{[I]}}$ refers to $S^{(S)}_T$ in \cite{LineSight} while the second-order source function $S_{\text{2ND}}$ can be expanded into its linear and quadratic parts
$S_{\text{2ND}}=S^{\text{[II]}} + S^{\text{[I,I]}}$.
The linear part $S^{\text{[II]}}$ refers to eq.~(40) of \cite{WayneHu} with the first-order perturbations replaced by the kernels of the corresponding second-order perturbations\footnote{The kernels $\mathcal{A}^{\text{[II]}}(\bold{k}_1,\bold{k}_2,\tau)$ are defined as \\$
\mathcal{A}^{\text{[II]}}(\bold{k},\tau)\equiv\int d\bold{k}_1d\bold{k}_2
(2\pi)^{-3/2}\delta(\bold{k}-\bold{k}_1-\bold{k}_2)
\mathcal{A}^{\text{[II]}}(\bold{k}_1,\bold{k}_2,\tau).$}. Analogous to well-known linear perturbation theory, $S^{\text{[II]}}$ contains contributions from intrinsic photon density, Sachs-Wolfe(SW), Doppler, integrated Sachs-Wolfe(ISW) effects of purely 2nd-order perturbations which require the full set of solutions from the 2nd-order Boltzmann\cite{Pitrou2nd} and Einstein Field equations\cite{2NDEFEs} (BE and EFE respectively). 

In contrast, the quadratic part $S^{\text{[I,I]}}$ requires solutions of 1st-order perturbations only and can be read in configuration space as\footnote{For simplicity, the notation for 1st-order perturbations is omitted.}
\begin{eqnarray}\label{QuadSc0}
S^{\text{[I,I]}} &=& 2e^{-\bar{\tau}}[S^{ij}\partial_j(\Phi+\Psi)\frac{\partial \triangle}{\partial n^i}-(\Phi+\Psi)n^i\partial_i\triangle\nonumber\\
&+&4(-n^i\partial_i\Phi+\Psi ')\triangle+(\Phi-\Psi)n^i\partial_i\Phi\nonumber\\
&+&2\Psi\Psi '+\mathfrak{C}^{\text{[I,I]}}]
\end{eqnarray}
where $S^{ij}$ is the screen projector \cite{Pitrou2nd}, $4\triangle\equiv\delta I/\bar{I}$ is the fractional brightness, $n^i$'s are the components of the observational direction, $\mathfrak{C}^{\text{[I,I]}}$ is the quadratic collision operator, and $\bar{\tau}$ is the optical thickness.\footnote{The quadratic collision $\mathfrak{C}^{\text{[I,I]}}$ contains only quadratic terms in 1st-order perturbaions and refers to eq.~(A.55) of \cite{PitrouBispec} in Fourier space.}

We rewrite the 1st term in the 2nd line  of Eq.~\eqref{QuadSc0} $\triangle n^i\partial_i\Phi=n^i\partial_i(\triangle\Phi)-\Phi n^i\partial_i\triangle$ and perform an integration by part for the 1st term of r.h.s., to obtain\footnote{Another integration by part is needed to transform $(\Psi ' -n^i\partial_i\Phi)\Phi$ into $\Phi(\Phi ' + \Psi ')$ (SW-ISW) and $\bar{\tau}'\Phi^2$ (SW-SW).}
\begin{eqnarray}\label{QuadSc}
S^{\text{[I,I]}} &=& 2e^{-\bar{\tau}}[S^{ij}\partial_j(\Phi+\Psi)\frac{\partial \triangle}{\partial n^i}-(\Phi+\Psi)n^i\partial_i\triangle\nonumber\\
&+&4(\Phi '+\Psi ')(\triangle+\Phi)+(\Phi-\Psi)n^i\partial_i\Phi+2\Psi\Psi ']\nonumber\\
&+&2g[(4\triangle+\tilde{\mathfrak{C}})\Phi+2\Phi^2+\tilde{\mathfrak{C}}^{\text{[I,I]}}]
\end{eqnarray}
where $g\equiv\bar{\tau}'e^{\bar{\tau}}$ and $\mathfrak{C}\equiv\bar{\tau}'\tilde{\mathfrak{C}}$.

In particular, the first and second term in the first line of eq.~\eqref{QuadSc} are the lensing and time-delay effects respectively. The term $\triangle(\Phi '+\Psi ')$ can be identified as the photon-ISW coupling when the perturbed photons are redshifted by the time-varying gravitational potential. Similarly, we can interpret $\Phi(\Phi '+\Psi ')$, $(4\triangle+\tilde{\mathfrak{C}})\Phi$ and $\Phi^2$ as the SW-ISW, photon-SW and SW-SW effects respectively. The term $(\Phi-\Psi)n^i\partial_i\Phi+2\Psi\Psi '$ is the quadratic part of the evolution equation of photon energy $p$ in 2nd-order, i.e.~$(dp/d\eta)^\text{[I,I]}$. We will discuss the contributions of these effects on the bispectrum later.

Our ultimate goal is to calculate the bispectrum as observed \emph{today}, which means that all the effects post recombination till today are ideally included in the calculation. However, an appropriate milestone towards this goal is the computation of the bispectrum \emph{at the end of recombination} which is what we undertake in this paper. This means that we cut off the numerical integration at this point.  

However, there exist source terms in this calculation which, via integrations by parts, are either zeroes or can be rescaled away \emph{today}, i.e. we can drop the boundary terms if they are taken today. In moving from eq.~\eqref{QuadSc0} to eq.~\eqref{QuadSc}, as we mentioned, we have performed integration by parts for the term $n^i\partial_i(\triangle\Phi)$ in the line of sight integral and dropped the boundary term evaluated today.
This means that we have secretly included post-recombination effects as a bonus.  More importantly, these terms turn out to be much easier numerically to deal with once integration by parts has been performed.


Thus, it is necessary to clarify which terms of the source function are considered when we restrict the time integration around the recombination. In our study, we include all the terms of the source function $S_\text{2ND}$ in eq.~\eqref{QuadSc} coupled with the function $g$ as it behaves like a dirac-delta function at recombination. We will also include all ISW-related effect at early-time\footnote{We also take $(dp/d\eta)^\text{[I,I]}$ into account but its contribution to the bispectrum is small.}, such as the purely 2nd-order ISW and photon-ISW coupling.\footnote{Effectively, we ignore the ISW effect due to late-time Dark Energy driven acceleration. We perform the time integration from $\tau=230$ to $1050$ Mpc to make sure that the early-time ISW effects are included.}. We ignore contributions from lensing, time-delay and late-time ISW-related effects.  The lensing and time-delay effects have been studied elsewhere \cite{LewisLensing,WHuTimedelay}.  Finally, we leave for a future publication the effects from the 2nd-order vector and tensor perturbations and the late-time ISW-related effects -- note that the former contributes throughout the entire time integration of the line-of-sight approach although they should be fairly small.



\smallskip
\emph{\textbf{Numerical Implementation}.}
To calculate the bispectrum in eq.~\eqref{Bispectrum}, we solve the perturbations numerically for both 1st and 2nd-order terms.  For the latter, we compute the kernels of the 2nd-order perturbations as functions of $k_1$, $k_2$, $\bold{k}_1\cdotp\bold{k}_1$ and $\tau$.

\begin{figure}[h]
\centering
\includegraphics[scale=0.5]{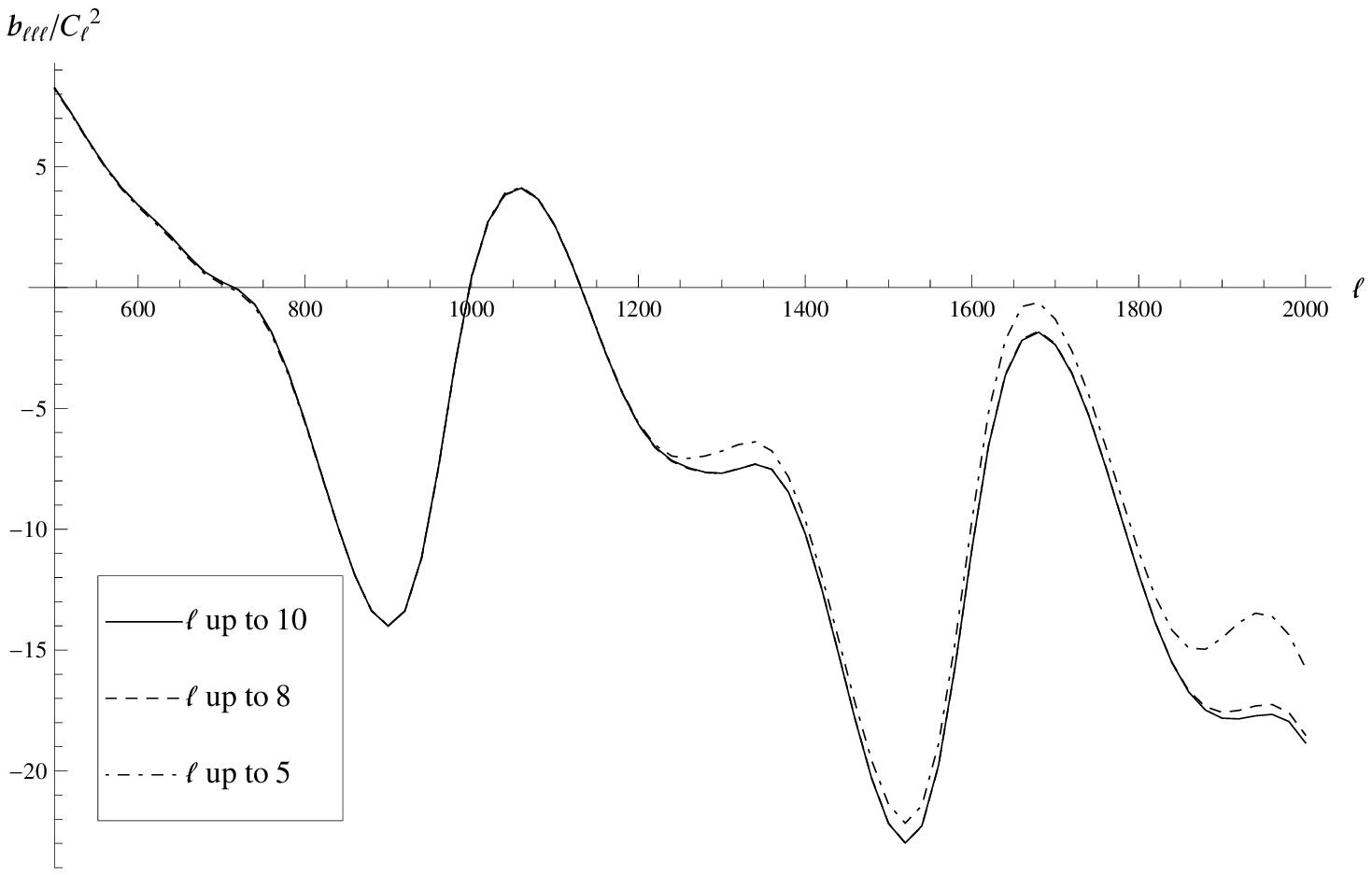}
\includegraphics[scale=0.5]{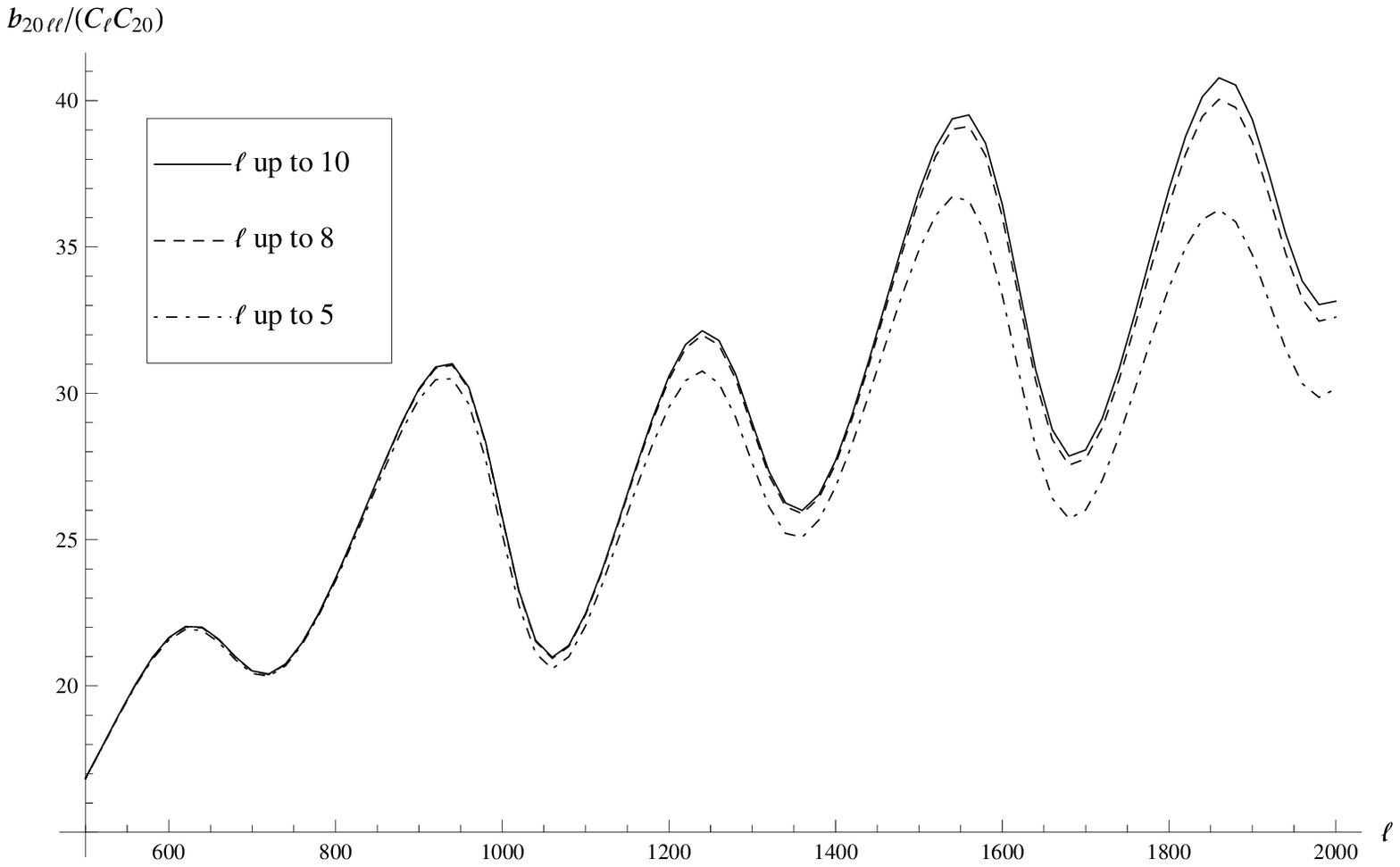}
\caption{The graphs of the bispectra generated from the quadratic source function $S^{\text{[I,I]}}$ against $l$ for equilateral(upper) and squeezed(lower) limit. The curves correspond to different $l$ truncations of the multipoles $S^{\text{[I,I]}}_{lm}$. We can see that the convergence occurs when $l$ goes up to 10.}
\label{ConvergeL}
\end{figure}

The solved 2nd-order perturbations can then be fed into the purely 2nd-order source function $S^{\text{[II]}}$. Although only 1st-order perturbations are needed to compute the quadratic source function $S^{\text{[I,I]}}$, it has to be decomposed into multipoles $S^{\text{[I,I]}}_{l,m}$. In principle, to close the hierarchy requires multipoles up to infinite $l$, but this is numerically consuming, and hence the usual prescription is to truncate the hierarchy once convergence is reached. For the bispectrum at recombination, we found that $l$ up to $10$ is sufficient (See Fig.~\ref{ConvergeL}).  We include all the $m\neq 0$ modes consistently up to $l=10$.

Our bispectrum at the squeezed limit is in good agreement with the analytical estimate in \cite{AnalyticSqu,AnalyticSqu2}. However, as we have indicated previously, our integration by part scheme has included different post recombination effects and hence we do not expect a full agreement.

In addition, we remark that the numerical accuracy of the multipoles $S^{\text{[I,I]}}_{l,m}$ with $m\neq 0$ cannot be checked with the analytic solution as they contribute negligibly in the squeezed limit. The reason is as follows. In this limit, $\bold{k}^\text{L}_1\ll \bold{k}^\text{S}_2,\bold{k}^\text{S}_3$, and combined with the conservation law of momenta, $\bold{k}^\text{S}_2$ is approximately parallel to $\bold{k}^\text{S}_3$ which is chosen to align with the $z$-axis of the multipole decomposition. Thus, modes with $m\neq 0$ are suppressed\footnote{This is true if we only consider scalar perturbations in 1st-order. In this case, we have only $m=0$ modes in 1st-order when $\bold{k}$ aligns with the $z$-axis of the multipole decomposition.}. Similarly, the 2nd-order vector and tensor perturbations are not important in the squeezed limit as they are sourced by $m=1$ and $m=2$ modes respectively.

\begin{figure}[h]
\centering
\includegraphics[scale=0.5]{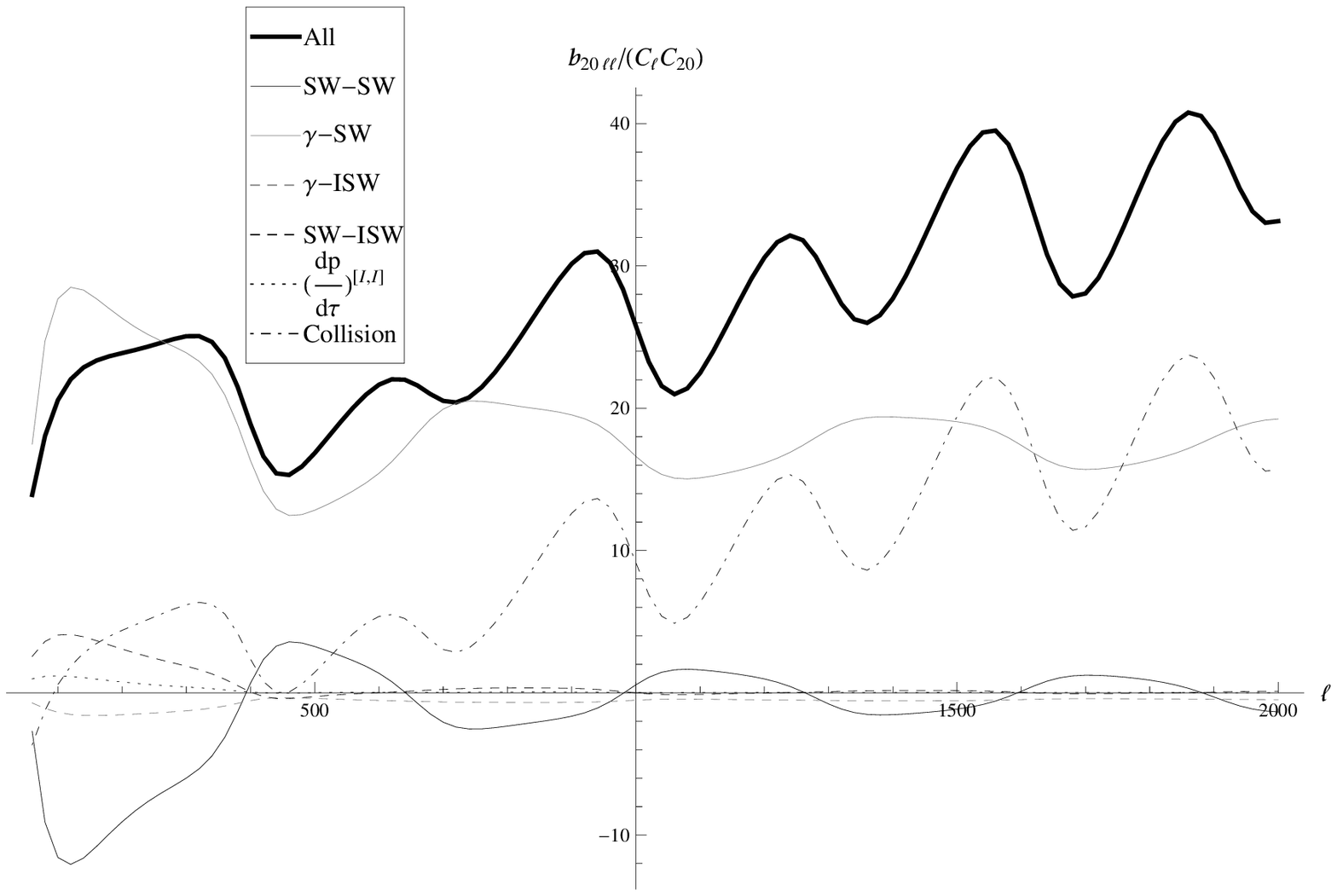}
\includegraphics[scale=0.5]{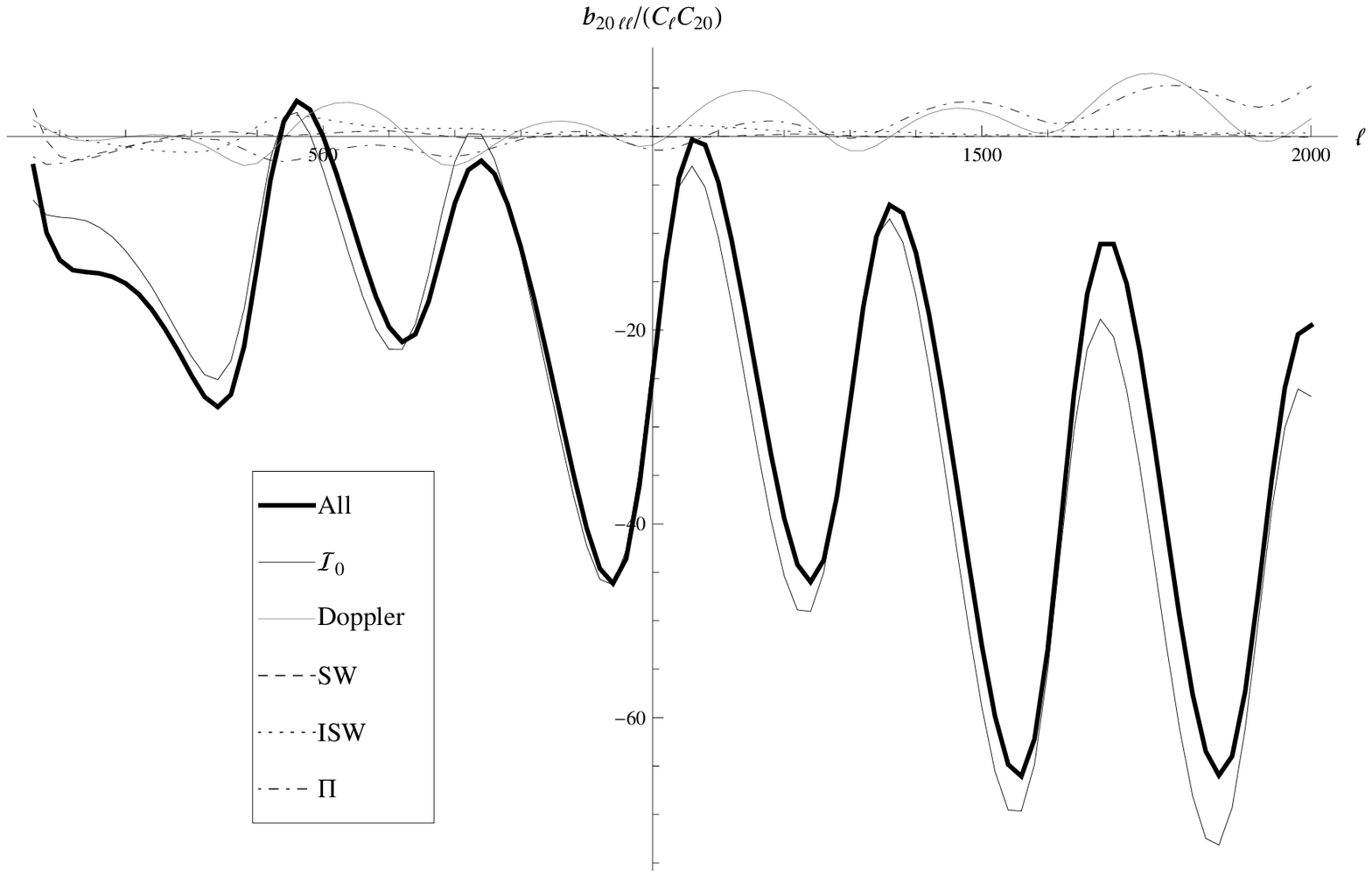}
\caption{The bispectra of the terms in the quadratic(upper) and linear(lower) source function in squeezed limit.}
\label{ScTerms}
\end{figure}

\smallskip
\emph{\textbf{Results and Discussion}.}
In Fig.~\ref{ScTerms}, we plot the bispectra of the terms in the source function $S_{\text{2ND}}$. We can see that the main contributions come from the effects of the Photon-SW, SW-SW and quadratic Collisions as well as the purely second-order SW, Doppler and anisotropic stress\footnote{Explicitly, the term $\Pi$ in \cite{LineSight}.} effects. The sum of the Photon-SW and the SW-SW effects has roughly the same value regardless of $l$. That is, the total bispectrum from these two effects is approximately proportional to the product of the power of the long and short wavelength modes, i.e. $C_{l_S} C_{l_L}$. This is because the power spectrum of the short wavelength mode comes mainly from intrinsic intensity, SW and Doppler effects while the power spectrum of the long wavelength mode is proportional to the square of the initial gravitational potential $\Phi$. This shifts the bispectrum from the purely second-order source terms up and suppresses the correlation between the bispectrum at recombination and that of the local type.

\begin{figure}[h]
\centering
\includegraphics[scale=0.6]{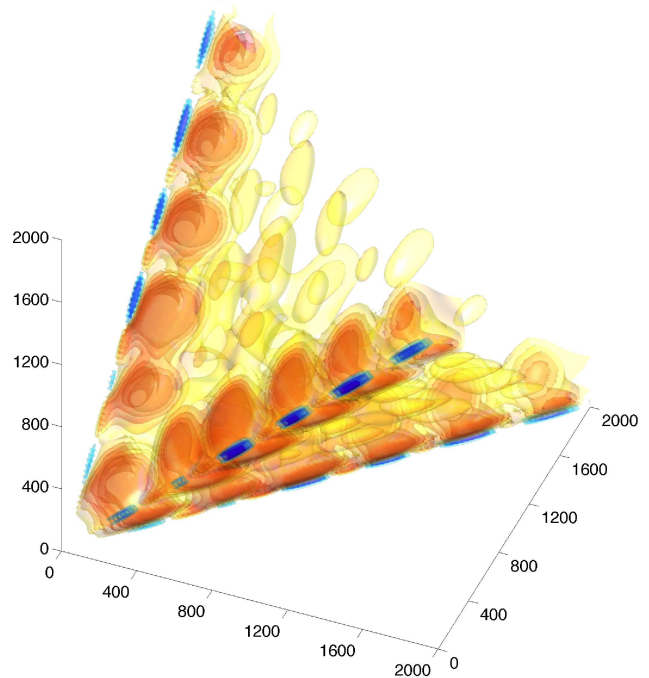}
\caption{The 3D plot of the reduced bispectrum generated at recombination. The bispectrum is normalized by the coefficient $D(l_1,l_2,l_3)$ defined in \cite{James} to remove an overall $l^{-4}$ scaling. We can see that the main contribution occurs in a fairly diffuse region near the edges of the tetrahedron, i.e.\ towards the squeezed limit. The red regions represent positive values while the blue regions represent negative values of the recombination bispectrum -- the oscillatory nature of the bispectrum means that it does not conform well to known templates such as the local or equilateral types. The signal to noise for the bispectrum is $0.69$ for $l_{\rm max} =2000$ as plotted.}
\label{Sec3D}
\end{figure}

\begin{figure}[h]
\centering
\includegraphics[scale=0.6]{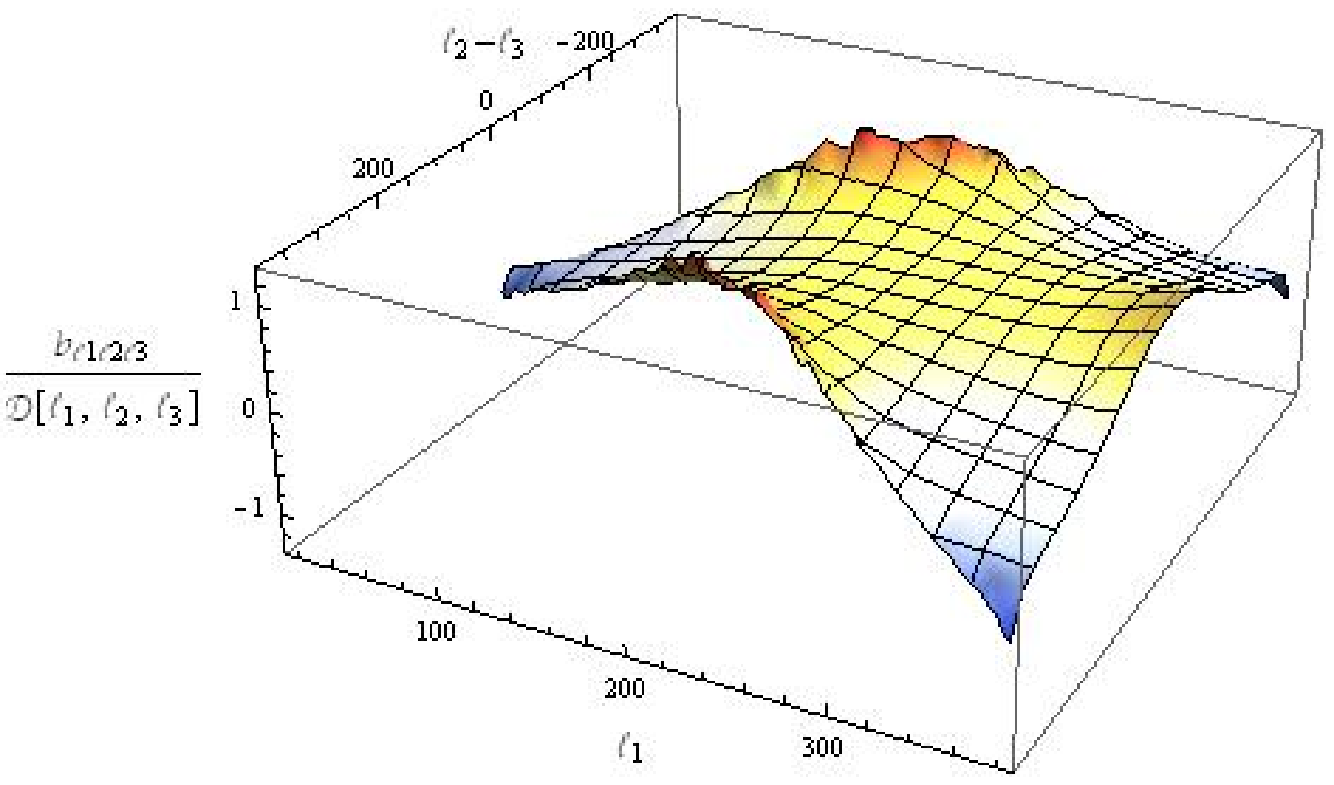}
\includegraphics[scale=0.6]{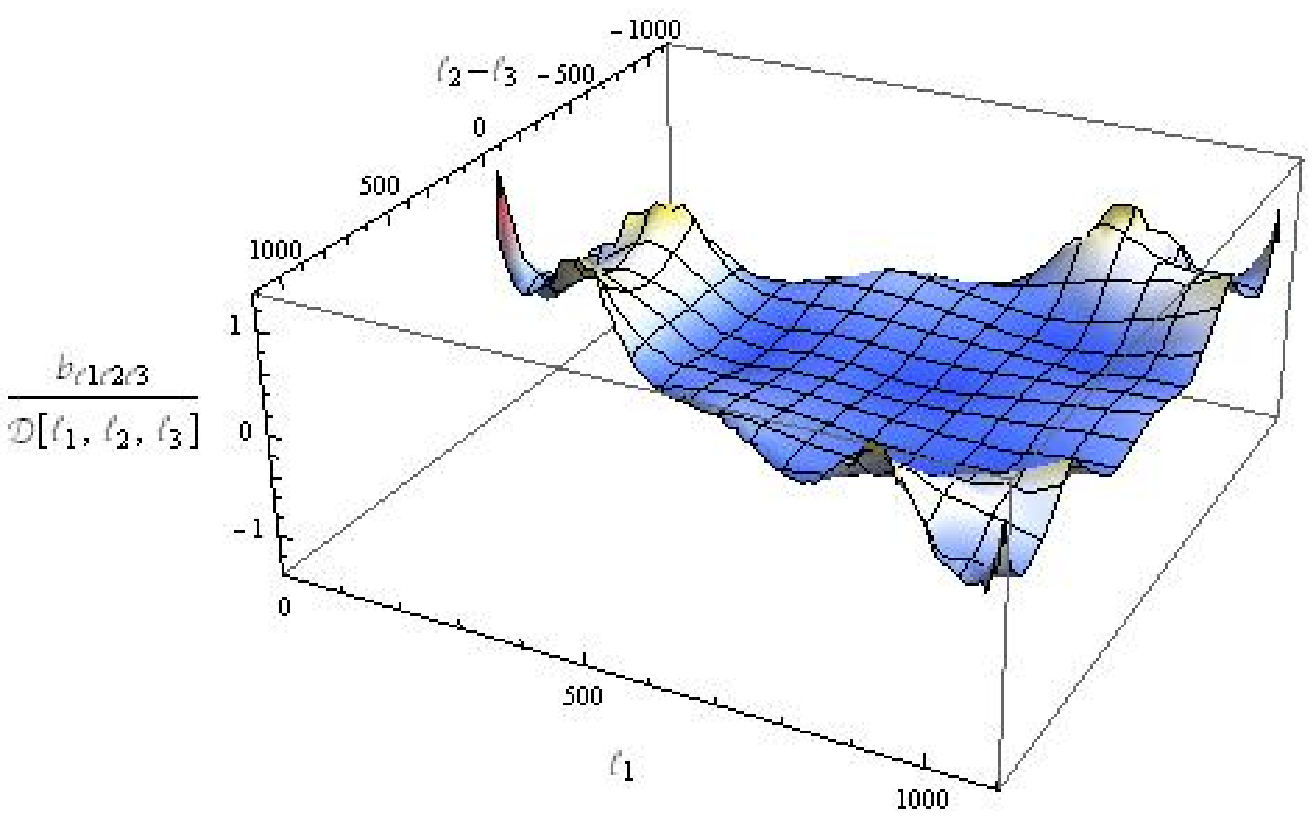}
\includegraphics[scale=0.6]{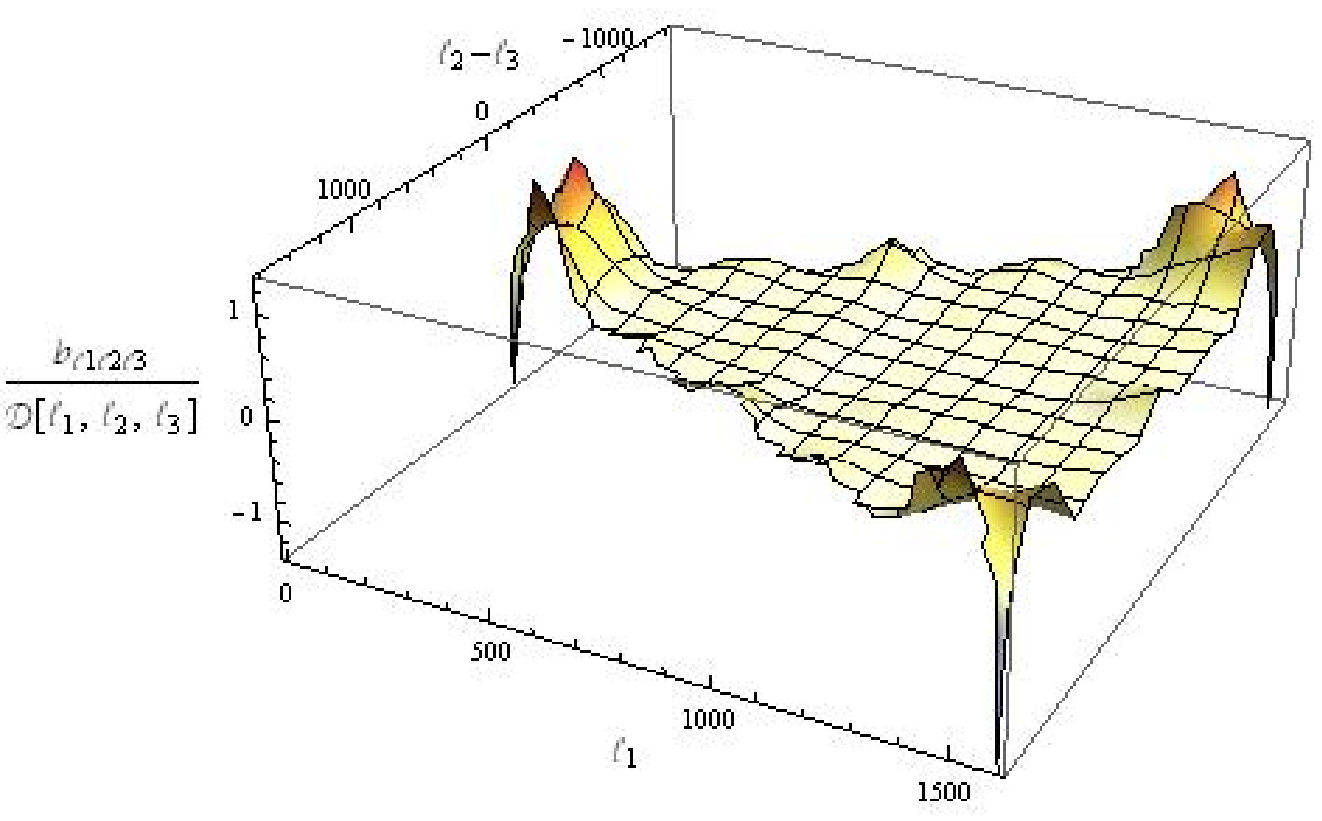}
\caption{The cross sections of the normalized reduced bispectrum in Fig.~\ref{Sec3D} with the conditions ${\textstyle\frac{1}{2}}(l_1+l_2+l_3)=400$ (upper panel), $1100$ (middle panel) and $1600$ (lower panel). They correspond to hyperslices through Fig. \ref{Sec3D}.}
\label{Sec2Dlow}
\end{figure}

We present the full bispectrum at recombination in Fig.~\ref{Sec3D}, showing isosurfaces of the bispectrum density. Although the main contribution to the recombination bispectrum is concentrated towards the edges of the tetrahedron, in the strongly squeezed limit the bispectrum fluctuates around zero and hence its correlation with the local type template is somewhat suppressed. What is clear from the figure is that the bispectrum does not correlate particularly well with the popular templates -- local, equilateral or orthogonal.   It possesses its own distinct shape with most power located between the local and equilateral limits. In Fig.~\ref{Sec2Dlow}, we show different tetrahedral cross-sections through the full recombination bispectrum taken at different summations $l_1+l_2+l_3 = $ const.   On the other hand, it contains features in the squeezed limit and along the edges  which reflect those appearing in the ISW-lensing bispectrum, but we will report on ISW cross-correlations in our longer paper \cite{SLS}.   

In Table \ref{TabFnl}, we summarize the effective $f_{\text{NL}}$'s, the normalized $F_{\text{NL}}$'s \cite{FNLDef} and their signal-to-noise ratios. The local-type $F_{\text{NL}}^{local}$ is small because of the fluctuating bispectrum in squeezed limit as we explained previously. The equilateral-type $F_{\text{NL}}^{equil}$ is small due to the lack of support in the interior of the bispectrum.
\begin{table}[htb]
 \caption{\label{TabFnl} The table of the effective $f_{\text{NL}}$'s, $F_{\text{NL}}$'s and signal-to-noise ratio $S/N$ of the local and equilateral templates correlated to the recombination bispectrum,  as well as its total signal (auto-correlation).  For ease of comparison, the $F_{\text{NL}}$ quantity normalises the integrated bispectrum signal for any shape relative to the $f_{\rm NL}=1$ local model.   It shows that the recombination bispectrum does not correlate well with local and equilateral templates. We have used $l_{max}=2000$ throughout.}
 \smallskip
 
 \begin{tabular}{lclcl}
  \hline
  \hline
	Model~~~~~~~~ & ~~~~~~~~$f_{\text{NL}}$~~~~~~~~ & $F_{\text{NL}}$~~~~~~ & $(S/N)$\\
  \hline
	Equilateral &	5.11	&	0.66	&	0.028 \\
	Local			&	0.88	&	0.88	&	0.22 \\
	Auto-correlation &--- &	3.19 & 0.69 \\
  \hline
  \hline
 \end{tabular}
 
\end{table}

These results are consistent in the squeezed limit with  \cite{AnalyticSqu,AnalyticSqu2}, confirming the correction of the  value $\sim 5$ as first calculated in \cite{PitrouBispec}.  In this work, however, we have focussed on increasing quantitative accuracy across the full tetrahedral domain beyond the squeezed limit.   We have computed the equilateral type  $f_{\text{NL}}^{\rm equil}= 5.11$ ($F_{\text{NL}} = 0.66$) which is modest but consistent with \cite{EquilFNL}.   However, the full  signal-to-noise ratio of the recombination bispectrum in our calculation out to $l_{max}=2000$ is substantial $(S/N)_{rec}\sim 0.69$ ($F_{\text{NL}} = 3.19$)  and larger than anticipated in the recent literature.   Confirmation of the magnitude of this signal will be important for the a quantitative analysis of the Planck data where this contribution should be incorporated in debiasing local and equilateral signals and in determining whether there is an overall primordial non-Gaussian signature in the data.   The recombination bispectrum will combine with the ISW-lensing bispectrum at about the 10\% level (or greater) and its correlation can affect the significance of this determination in the Planck data (recall that ISW can bias the local signal by as much as $f_{\rm NL} = 9.5$ \cite{LewisLensing}).  For this reason we shall continue to incorporate more physical effects in our numerical pipeline to improve this quantitative analysis further. 

\smallskip
\emph{\textbf{Summary}.}
In this paper, we focus on the bispectrum generated at recombination across the full range of multipole combinations.  We find that the effective $f_{\text{NL}}$'s of the equilateral and local types are $5.11$ and $0.88$ respectively, while the overall signal-to-noise is $(S/N)_{\rm rec}=0.69$, all calculated using $l_{max}=2000$. We note from Fig.~\ref{Sec3D} that the bispectrum possesses its own distinct features differentiating it from well-known templates, such as local, equilateral and ISW-lensing.  To complete the full calculation of this late-time bispectrum will require the inclusion of the time-delay and lensing effects up to second-order (both which have been studied separately in the literature), the addition of the second-order vector and tensor perturbations, and finally the late-time ISW effects. These will be addressed in a future publication \cite{SLS}.  With improving precision, this recombination bispectrum should be included in the analysis of future CMB experiments.   

\begin{acknowledgments}
\emph{Acknowledgements}. 
We are grateful for many informative conversations  with Julien Lesgourgues, T. Tram, Christian Fidler, Zhiqi Huang and Christophe Pitrou (including comparisons with CMBQUICK\cite{CMBQuick}). We would also like to thank James Fergusson for  many useful discussions and help with bispectrum visualisation.  The numerical simulations were implemented and performed on the COSMOS supercomputer, part of the DiRAC HPC Facility which is funded by STFC and BIS. We thank Andrey Kaliazin for computational support and technical advice.   E. A. Lim, S. Su and E.P.S. Shellard were supported by STFC grant ST/F002998/1 and the Centre for Theoretical Cosmology. The numerical code was developed based on CAMB \cite{CAMB}.
\end{acknowledgments}

\medskip

\noindent \emph{Note added in proof:} While this paper was being prepared, a preprint appeared with an alternative numerical implementation which also tackles second-order CMB perturbations \cite{Zhiqi}.   This short paper offers a number of tests in the squeezed limit including a forecast $f_{NL}^{local}\approx 0.82$, in agreement with our results.   An equilateral prediction was not given but the overall signal  quoted  \cite{Zhiqi} $(S/N)_{rec} = 0.47$ is lower than our result. This discrepancy may reflect the range of physical effects which have been incorporated in the two approaches.   It is unlikely to be due to tensor and vector contributions which are subdominant, but it may reflect a different hierarchy truncation scheme of the $S^{\text{[I,I]}}_{l,m}$ multipoles or the integration by parts approach they adopt which seems to include some different post recombination effects.


\smallskip

\begin{thebibliography}{99}
\bibitem{NGReview} Bartolo N., Matarrese S., Riotto A., Adv. Astron. {\bf 2010}, 157079 (2010).
\bibitem{TempDef} Nitta D. et al., J. Cosmol. Astropart. Phys. {\bf 05}, 014 (2009).
\bibitem{NBispec2} Senatorea L., Tassevc S., Zaldarriaga M., J. Cosmol. Astropart. Phys. {\bf 09}, 038 (2009).
\bibitem{PitrouBispec} Pitrou C., Uzan J.-P., Bernardeau F., J. Cosmol. Astropart. Phys. {\bf 07}, 003 (2010).
\bibitem{EquilFNL} Bartolo N., Riotto A., J. Cosmol. Astropart. Phys. {\bf 03}, 017 (2009).
\bibitem{1stOpacity} Novosyadlyj B., Mon. Not. Roy. Astron. Soc. {\bf 370}, 1771 (2006).
\bibitem{LineSight} Seljak U., Zaldarriaga M., Astrophys. J. {\bf 469}, 437 (1996).
\bibitem{WayneHu} Hu W. et al., Phys. Rev. D {\bf 57}, 6 (1998).
\bibitem{Pitrou2nd} Pitrou C., Class. Quant. Grav. {\bf 26}, 065006 (2009).
\bibitem{2NDEFEs}  Bartolo N., Matarrese S., Riotto A., J. Cosmol. Astropart. Phys. {\bf 06}, 024 (2006).
\bibitem{LewisLensing} Lewis A., Challinor A., Hanson D., J. Cosmol. Astropart. Phys. {\bf 03}, 018(2011).
\bibitem{WHuTimedelay} Hu W., Cooray A., Phys. Rev. D {\bf 63}, 023504 (2001).
\bibitem{AnalyticSqu} Creminelli P., Pitrou C., Vernizzi F., J. Cosmol. Astropart. Phys. {\bf 11}, 025 (2011).
\bibitem{AnalyticSqu2} Bartolo N., Matarrese S., Riotto A., J. Cosmol. Astropart. Phys. {\bf 1202} 017 (2012).
\bibitem{FlatSky} Bernardeau F., Pitrou C., Uzan J.-P., J. Cosmol. Astropart. Phys. {\bf 02}, 015 (2011).
\bibitem{James} Fergusson J. R., Shellard E. P. S., Phys. Rev. D {\bf 80}, 043510 (2009).
\bibitem{FNLDef} Fergusson J. R., Liguori M., Shellard E. P. S. (2010), arXiv:1006.1642.
\bibitem{SLS} Su, S, Lim, E., Shellard, E.P.S., in preparation (2013).
\bibitem{CMBQuick} http://www2.iap.fr/users/pitrou/cmbquick.htm.
\bibitem{Zhiqi} Huang Z., Vernizzi F. (2012), 
arXiv:1212.3573.
\bibitem{CAMB} Lewis A., Challinor A., Lasenby A., Astrophys. J. {\bf 538}, 473 (2000).
\end{thebibliography}
\end{document}